\documentclass[twocolumn]{aastex7}   
\newcommand{\feh}{[\rm{Fe/H}]}
\newcommand{\teff}{$T_{\rm{eff}}$}
\newcommand{\kms}{km~s$^{-1}$}
\usepackage{xcolor}	
\usepackage{graphicx}	
\usepackage{float}

\begin{document} 

\title{Visual Orbits of Spectroscopic Binaries with the CHARA Array. V. \\ HD~210763 and HD~221950}

% confirmed co-authors
\author[0000-0002-9903-9911]{Kathryn V. Lester} 
\affil{Mount Holyoke College, South Hadley, MA, USA}
\affil{Planetary Science Institute, Tucson, AZ, USA}
\email{klester@psi.edu}

\author[0009-0000-8159-7491]{Josephine Singleton}  
\affil{Mount Holyoke College, South Hadley, MA, USA}
\email{singl22j@mtholyoke.edu }

\author{Alexandra M. Twomey}    
\affil{Mount Holyoke College, South Hadley, MA, USA}
\email{twome22a@mtholyoke.edu }

\author{Viktoria Sargent}     
\affil{Mount Holyoke College, South Hadley, MA, USA}
\email{sarge24v@mtholyoke.edu}

\author[0009-0004-3037-0448]{Sally V. Shepherd}    
\affil{Mount Holyoke College, South Hadley, MA, USA}
\email{sheph26s@mtholyoke.edu}

\author[0000-0001-8537-3583]{Douglas R. Gies} 
\affil{Center for High Angular Resolution Astronomy and Department of Physics \& Astronomy, Georgia State University, Atlanta, GA, USA}
\email{ dgies@gsu.edu}

\author[0000-0001-5415-9189]{Gail H. Schaefer}    
\affil{The CHARA Array of Georgia State University, Mount Wilson Observatory, Mount Wilson, CA, USA}
\email{gschaefer@gsu.edu}

\author[0000-0001-9745-5834]{Cyprien Lanthermann}    
\affil{The CHARA Array of Georgia State University, Mount Wilson Observatory, Mount Wilson, CA, USA}
\email{clanthermann@gsu.edu}

\author[0000-0003-3045-5148]{Jeremy Jones}    
\affil{Center for High Angular Resolution Astronomy and Department of Physics \& Astronomy, Georgia State University, Atlanta, GA, USA}
\email{jjones176@gsu.edu}

\author[0000-0003-2075-5227]{Katherine Shepard}    
\affil{Department of Physics \& Astronomy, Vanderbilt University, TN, USA}
\email{katherine.shepard.1@vanderbilt.edu }

\author[0000-0002-2208-6541]{Narsireddy Anugu}      
\affil{The CHARA Array of Georgia State University, Mount Wilson Observatory, Mount Wilson, CA, USA}
\email{nanugu@gsu.edu}

\author[0000-0001-9764-2357]{Claire L. Davies}      
\affil{Astrophysics Group, Department of Physics \& Astronomy, University of Exeter, Exeter, UK}
\email{ C.Davies3@exeter.ac.uk}

\author[0000-0002-9061-2865]{Todd J. Henry} 
\affil{RECONS Institute, Chambersburg, PA, USA}
\email{thenry88@gsu.edu}

\author[0000-0003-4568-2079]{Hodari-Sadiki Hubbard-James}     
\affil{Department of Physics and Astronomy, Agnes Scott College, Decatur, GA, USA}
\email{hjames@agnesscott.edu}

\author[0000-0003-0193-2187]{Wei-Chun Jao} 
\affil{Department of Physics \& Astronomy, Georgia State University, Atlanta, GA, USA}
\email{ wjao@gsu.edu}

\author[0000-0001-6017-8773]{Stefan Kraus} 
\affil{Astrophysics Group, Department of Physics \& Astronomy, University of Exeter, Exeter, UK}
\email{S.Kraus@exeter.ac.uk}

\author[0000-0002-3380-3307]{John D. Monnier}  
\affil{Department of Astronomy, University of Michigan, Ann Arbor, MI, USA}
\email{monnier@umich.edu}

\author[0000-0003-1324-0495]{Leonardo A. Paredes}      
\affil{Steward Observatory \& Department of Astronomy, University of Arizona, AZ, USA}
\email{lparedes@arizona.edu}

\author[0000-0001-5980-0246]{Benjamin R. Setterholm} 
\affil{Max-Planck-Institut für Astronomie, Heidelberg, Germany}
\affil{Department of Astronomy, University of Michigan, Ann Arbor, MI, USA}
\email{ setterholm@mpia.de}

\correspondingauthor{Kathryn Lester}

% -----------------------------------------------------------------------------
\begin{abstract}

We present the visual orbits and dynamical masses of two longer period spectroscopic binary stars, HD~210763 and HD~221950, using long baseline interferometry with the CHARA Array and high resolution spectroscopy with the APO 3.5~m and CTIO 1.5~m telescopes. By combining the astrometric and radial velocity observations, we solve for the full, three-dimensional orbits and determine the stellar masses to within 0.5\% uncertainty and the distance to within 0.2\% uncertainty. For HD~210763, we found component masses of $M_1 = 1.748\pm0.008 M_\odot$, $M_2 = 1.492\pm0.006 M_\odot$. For HD~221950, we found masses of $M_1 = 1.098\pm0.006 M_\odot$, $M_2 = 1.031\pm0.005 M_\odot$. 
We then estimate the effective temperature and radius of each component star through disentangling and spectral energy distribution analyses. We compare the observed stellar parameters to the predictions of the stellar evolution models and estimate the system ages. The primary component of HD~210763 is at the end of the main sequence while the secondary component is on the main sequence, providing a tight age constraint for this system at  $1.6 \pm 0.1$ Gyr. Both components of HD~221950 are still on the main sequence with an age of $3.76 \pm0.38$ Gyr. These systems have longer orbital periods, beyond the tidal circularization limit, and therefore are better proxies for single stars and tests of stellar evolution models than short period, eclipsing systems. 

\end{abstract}


% -----------------------------------------------------------------------------
\section{Introduction}\label{section:intro}

Precise fundamental parameters of the stars in binary systems are essential for determining orbital demographics \citep{raghavan10, bordier22}, testing models of stellar structure and evolution \citep[e.g.,][]{claret18, morales22}, calibrating mass and distance determination methods for single stars \citep{torres10, chaplin13, gallenne18}, and studying how binary stars form and evolve \citep[e.g.,][]{richardson21}. However, systems with periods less than 10 days are subject to tidal forces that deform the stellar photospheres and cause the orbit to shrink and circularize, so these systems are less reliable tests of single star evolution. Precise parameters of binaries with periods greater than the tidal circularization period of roughly 10 days \citep{raghavan10} are needed to provide much better tests. Because these systems are less likely to be eclipsing, their orbital inclinations and stellar parameters must be obtained by fitting the visual orbits instead. 

We continue our paper series measuring the visual orbits of spectroscopic binary stars with long baseline interferometry \citep[Papers I--IV,][]{lester19a, lester19b, lester20, lester23} in order to precisely determine the component stars' masses and distances. Our sample contains intermediate mass main sequence stars (A and F type) in binary systems with orbital periods longer than 14 days. We present here the results for two new systems, HD~210763 and HD~221950, with periods four times longer than the tidal circularization period. 

HD~210763 (HR 8467, HIP 109647) contains a pair of F-type stars in a 42 day orbit. \citet{nadal83} published the first double-lined spectroscopic orbit for this system, and \citet{fekel11} and \citet{katoh13} both measured updated orbits using high resolution spectra. \citet{gallenne23} measured the first visual orbit of HD~210763 with VLTI; we confirm their results in this work. \citet{fekel11} also noted that the primary component of HD~210763 is at the beginning of the Hertzsprung gap, so this system will provide an important test of the sub-giant branch position in evolutionary models. 

HD~221950 (16 Psc, HR 8954, HIP 116495) contains a pair of F-type stars in a 45 day orbit. \citet{strobel74} measured the first double-lined orbit for this system, and \citet{tomkin08} measured an updated spectoscopic orbit with higher resolution observations. We present the first visual orbit of this system. Additionally, \citet{strobel68} noted that HD~221950 is metal deficient and \citet{tomkin08} estimated the iron abundance to be $\feh \approx -0.5$, which we used for our template spectra. 

Our spectroscopic observations and radial velocity measurement are presented in Section~\ref{section:spec}, and our interferometric observations and relative astrometry measurements are presented in Section~\ref{section:inter}. We describe the combined visual and spectroscopic orbital solution in Section~\ref{sect:vbsbfit}. The temperatures and radii of each component from spectral disentangling and SED analysis are presented in Section~\ref{sect:params}. We compare the results to stellar evolution models in Section~\ref{sect:evofit} to estimate the age of each system, then discuss our conclusions in Section~\ref{sect:discussion}.

% -----------------------------------------------------------------------------
\section{Spectroscopy}\label{section:spec}

We obtained high resolution spectra of our stars using two optical spectrographs: the APO 3.5~m telescope's ARC echelle spectrograph \citep[ARCES;][]{arces} and the CTIO 1.5m telescope's CHIRON echelle spectrograph \citep{chiron, paredes21}. ARCES covers 3500-10500 \AA\ over 107 orders at an average resolving power of $R\sim30,000$. We reduced this data using standard echelle procedures in IRAF, as described in the guide by Karen Kinemuchi\footnote{\href{http://astronomy.nmsu.edu/apo-wiki/lib/exe/fetch.php?media=wiki:arces:kinemuchi\_arces\_cookbook.pdf}{http://astronomy.nmsu.edu/apo-wiki/lib/exe/fetch.php?media=wiki:arces:kinemuchi\_arces\_cookbook.pdf}}, including bias subtraction, 1D flat field correction, and wavelength calibration using ThAr lamp spectra. CHIRON covers 4500-8800 \AA\ over 60 echelle orders. We observed HD~210763 in fiber mode at  $R\sim25,000$ and HD~221950 in slicer mode at $R\sim79,000$. These data were reduced with the CHIRON team's pipeline \citep{paredes21}. We then normalized both sets of spectra using the procedure in Appendix~A of \citet{kolbas15}, in which the blaze functions for orders with strong spectral lines were interpolated from the polynomial fits to nearby featureless orders.

\begin{deluxetable}{lcrrc}	
\tablewidth{0pt}
\tabletypesize{\footnotesize}
\tablecaption{  Radial Velocity Measurements for HD 210763 \vspace{6pt} \label{rvtable1}    }
\tablehead{ 
\colhead{UT Date} & \colhead{MJD} & \colhead{RV$_1$ (km~s$^{-1}$)} & \colhead{RV$_2$ (km~s$^{-1}$)}& \colhead{Instrument}  
}
\startdata
%    DATE	MJD		RV1		ERV1		RV2		ERV2		Source
\hline
2016-10-15   &	57676.2113	&  $ 36.00 \pm 0.25$  &  $-10.67 \pm 1.11$   & ARCES  \\
2016-11-19   &	57711.1171	&  $ 75.61 \pm 0.48$  &  $-57.52 \pm 1.32$   & ARCES  \\
2016-12-15   &	57737.0365	&  $ -6.60 \pm 0.33$  &  $ 41.11 \pm 0.60$   & ARCES  \\
2017-09-02   &	57998.2356	&  $-19.32 \pm 0.59$  &  $ 55.15 \pm 0.95$   & ARCES  \\
2018-12-24   &	58476.0513	&  $ 62.24 \pm 0.24$  &  $-41.64 \pm 0.69$   & ARCES  \\
2019-07-08   &	58672.4874	&  $-12.57 \pm 0.42$  &  $ 46.23 \pm 1.03$   & ARCES  \\
2019-08-31   &	58726.1714	&  $ 56.17 \pm 0.29$  &  $-32.98 \pm 0.90$   & CHIRON \\
2019-09-01   &	58727.2239	&  $ 76.15 \pm 0.45$  &  $-58.64 \pm 0.95$   & CHIRON \\
2019-09-02   &	58728.2286	&  $ 76.05 \pm 0.52$  &  $-57.97 \pm 1.29$   & CHIRON \\
2019-09-03   &	58729.1518	&  $ 70.53 \pm 0.42$  &  $-50.63 \pm 2.05$   & CHIRON \\
2019-09-04   &	58730.1455	&  $ 63.25 \pm 0.29$  &  $-41.31 \pm 1.59$   & CHIRON \\
2019-09-05   &	58731.1574	&  $ 57.42 \pm 0.25$  &  $-34.22 \pm 0.76$   & CHIRON \\
2019-10-05   &	58761.1424	&  $-19.54 \pm 0.72$  &  $ 55.45 \pm 0.80$   & CHIRON \\
2019-10-07   &	58763.0150	&  $-22.80 \pm 0.43$  &  $ 58.03 \pm 1.63$   & CHIRON \\
2019-10-08   &	58764.0492	&  $-22.62 \pm 0.59$  &  $ 57.99 \pm 0.63$   & CHIRON \\
2019-10-09   &	58765.0401	&  $-21.22 \pm 0.68$  &  $ 55.79 \pm 0.99$   & CHIRON \\
2019-10-10   &	58766.0338	&  $-14.29 \pm 0.42$  &  $ 49.97 \pm 0.55$   & CHIRON \\
2019-10-12   &	58768.0948	&  $ 39.62 \pm 0.33$  &  $-13.66 \pm 1.49$   & CHIRON \\
2019-10-13   &	58769.1598	&  $ 71.07 \pm 0.42$  &  $-51.04 \pm 2.11$   & CHIRON \\
2019-10-15   &	58771.1380	&  $ 73.75 \pm 0.63$  &  $-53.59 \pm 2.12$   & CHIRON \\
2019-10-18   &	58774.1018	&  $ 53.83 \pm 0.21$  &  $-31.01 \pm 0.70$   & ARCES  \\
2019-11-11   &	58798.0533	&  $-10.10 \pm 0.20$  &  $ 44.22 \pm 0.52$   & CHIRON \\
2019-11-13   &	58800.0740	&  $-13.61 \pm 0.45$  &  $ 49.13 \pm 0.83$   & CHIRON \\
2019-11-15   &	58802.0701	&  $-17.68 \pm 0.29$  &  $ 52.53 \pm 0.91$   & CHIRON \\
2019-11-17   &	58804.0461	&  $-20.74 \pm 0.71$  &  $ 56.32 \pm 1.18$   & CHIRON \\
2020-06-07   &	59007.4427	&  $ -5.93 \pm 0.33$  &  $ 37.43 \pm 0.93$   & ARCES  \\
2020-06-15   &	59015.3685	&  $-19.78 \pm 0.34$  &  $ 54.73 \pm 0.70$   & ARCES  \\
2020-06-29   &	59029.3650	&  $ 48.41 \pm 0.29$  &  $-25.22 \pm 1.16$   & ARCES  \\
\enddata
\end{deluxetable}

\begin{deluxetable}{lcrrc}	
\tablewidth{0pt}
\tabletypesize{\footnotesize}
\tablecaption{  Radial Velocity Measurements for HD 221950\vspace{6pt} \label{rvtable2}    }
\tablehead{ 
\colhead{UT Date} & \colhead{MJD} & \colhead{RV$_1$ (km~s$^{-1}$)} & \colhead{RV$_2$ (km~s$^{-1}$)}& \colhead{Instrument}  
}
\startdata
%    DATE	MJD		RV1		ERV1		RV2		ERV2		Source
\hline
2016-09-14   &	57645.2576	&  $ 9.60 \pm 0.92$  &  $94.95 \pm 0.64$  &  ARCES \\
2019-09-08   &	58734.2671	&  $21.94 \pm 0.19$  &  $78.78 \pm 0.22$  & CHIRON \\
2019-09-12   &	58738.2318	&  $ 6.01 \pm 0.18$  &  $95.65 \pm 0.08$  & CHIRON \\
2019-09-13   &	58739.1328	&  $ 7.11 \pm 0.07$  &  $94.51 \pm 0.09$  & CHIRON \\
2019-09-14   &	58740.1580	&  $ 9.36 \pm 0.11$  &  $92.12 \pm 0.12$  & CHIRON \\
2019-09-15   &	58741.1337	&  $12.05 \pm 0.10$  &  $89.41 \pm 0.11$  & CHIRON \\
2019-09-16   &	58742.1371	&  $15.12 \pm 0.15$  &  $86.38 \pm 0.22$  & CHIRON \\
2019-09-17   &	58743.1614	&  $18.13 \pm 0.31$  &  $83.13 \pm 0.21$  & CHIRON \\
2019-09-18   &	58744.1459	&  $21.18 \pm 0.17$  &  $79.56 \pm 0.30$  & CHIRON \\
2019-09-19   &	58745.1297	&  $24.35 \pm 0.13$  &  $76.31 \pm 0.08$  & CHIRON \\
2019-09-20   &	58746.1773	&  $27.87 \pm 0.11$  &  $72.70 \pm 0.14$  & CHIRON \\
2019-09-21   &	58747.2000	&  $31.05 \pm 0.17$  &  $69.39 \pm 0.16$  & CHIRON \\
2019-10-13   &	58769.1668	&  $84.90 \pm 0.10$  &  $11.80 \pm 0.35$  & CHIRON \\
2019-10-15   &	58771.1695	&  $86.35 \pm 0.08$  &  $10.57 \pm 0.28$  & CHIRON \\
2019-10-16   &	58772.1183	&  $85.77 \pm 0.11$  &  $11.04 \pm 0.30$  & CHIRON \\
2019-10-17   &	58773.1161	&  $83.76 \pm 0.15$  &  $12.94 \pm 0.32$  & CHIRON \\
2019-10-17   &	58773.9905	&  $80.68 \pm 0.08$  &  $16.06 \pm 0.09$  & CHIRON \\
2019-10-25   &	58781.1019	&  $11.40 \pm 0.05$  &  $90.22 \pm 0.17$  & CHIRON \\
2020-06-07   &	59007.4337	&  $18.80 \pm 1.30$  &  $83.58 \pm 1.34$  &  ARCES \\
2020-08-26   &	59087.2247	&  $84.51 \pm 0.83$  &  $12.03 \pm 1.52$  &  ARCES \\
2021-05-24   &	59358.4611	&  $81.57 \pm 1.13$  &  $13.27 \pm 1.31$  &  ARCES \\
2022-06-11   &	59741.4386	&  $12.92 \pm 0.99$  &  $88.62 \pm 1.17$  &  ARCES \\
2022-10-04   &	59856.3722	&  $77.57 \pm 1.11$  &  $17.29 \pm 1.69$  &  ARCES \\
2023-01-22   &	59966.0844	&  $ 6.60 \pm 0.70$  &  $95.62 \pm 0.78$  &  ARCES \\
2023-01-30   &	59974.0703	&  $27.80 \pm 0.76$  &  $70.82 \pm 2.15$  &  ARCES \\
2023-06-03   &	60098.4828	&  $21.42 \pm 1.76$  &  $83.60 \pm 2.05$  &  ARCES \\
2023-06-11   &	60106.4238	&  $16.60 \pm 0.92$  &  $84.65 \pm 0.93$  &  ARCES \\
2023-10-02   &	60219.1847	&  $77.34 \pm 0.62$  &  $21.08 \pm 1.06$  &  ARCES \\
2023-10-06   &	60223.2631	&  $83.62 \pm 0.65$  &  $12.23 \pm 1.20$  &  ARCES \\
2023-10-31   &	60248.2506	&  $35.82 \pm 0.91$  &  $65.42 \pm 0.76$  &  ARCES \\
\enddata
\end{deluxetable}

Radial velocities (RVs) were measured with the multi-order TODCOR method \citep{todcor1, todcor2}. TODCOR calculates the cross correlation function (CCF) for a grid of primary and secondary radial velocities. Template spectra were created with PHOENIX model spectra \citep{phoenix} for atmospheric parameter estimates (\teff, $\log g, v\sin i$) from the literature. For HD 210763, \citet{fekel11} measured effective temperatures of $6270\pm200$ K and $6260\pm200$ K and projected rotational velocities of $10.6\pm1.0$ and $9.5\pm2.0$ \kms~ for the primary and secondary components. For HD~221950, \citet{tomkin08} measured an effective temperature of $6240\pm150$ K for both components and projected rotational velocities of $9.5\pm1.0$ and  $6.3\pm1.0$ \kms. The CCFs for each echelle order were added together to find the location of the maximum CCF, corresponding to the best-fit radial velocities ($RV_1$, $RV_2$) and their uncertainties ($\sigma_1$, $\sigma_2$).  TODCOR also calculates the monochromatic flux ratio ($f_2/f_1$) near 600~nm, which was 0.38 for HD~210763 and 0.77 for HD~221950.

We accounted for differences in the zero-point offset of each spectrograph by first fitting separate orbital solutions to each data set using a least-squares optimizer. We solved for the six spectroscopic orbital parameters of each system: the orbital period ($P$), epoch of periastron ($T$), eccentricity ($e$), longitude of periastron of the primary star ($\omega_1$), velocity semi-amplitudes ($K_1$, $K_2$), and systemic velocity ($\gamma$). We also fit orbits to the previously published velocities of HD~210763 from \citet{fekel11} and HD~221950 from \citet{tomkin08}, using uncertainties calculated from the weights reported in these papers ($\sigma = 1/\sqrt{weight}$). The ARCES and CHIRON RVs were then offset so their systemic velocity solutions matched those of the literature datasets, though the offsets were quite small (less than 0.4 \kms). Our final radial velocities are listed in Table~\ref{rvtable1} for HD~210763 and Table~\ref{rvtable2} for HD~221950, with the UT date, modified Julian date (MJD), heliocentric radial velocity and uncertainty of each component, and the instrument used.

% -----------------------------------------------------------------------------
\section{Interferometry}\label{section:inter}

We observed both binaries with the CHARA Array interferometer using the MIRC-X \citep{mircx} and MYSTIC \citep{mystic} beam combiners. CHARA has six 1.0~m telescopes arranged in a Y-shape with baselines ranging from 34--331~m \citep{chara}. MIRC-X  and MYSTIC combine the $H$- and $K$-band light, respectively, from up to six telescopes and disperses the light into eight narrowband spectral channels at $R=50$. Our observations are listed in Tables~\ref{relpos1} for HD~210763 and \ref{relpos2} for HD~221950. The 2020 data used only MIRC-X, while the 2024 data used both MIRC-X and MYSTIC. All data were reduced using the pipeline developed by the MIRC-X team\footnote{\href{https://gitlab.chara.gsu.edu/lebouquj/mircx_pipeline}{https://gitlab.chara.gsu.edu/lebouquj/mircx\_pipeline}, v1.3.5 for the 2020 data and v1.5.0 for the 2024 data} to determine squared visibilities ($V^2$) for each baseline and closure phases (CP) for each closed triangle. We corrected for any instrumental and atmospheric effects using observations of calibrators stars, for which uniform disk angular diameters have been previously measured and listed on SearchCal \citep{searchcal}:
HD 207095 with $\theta_H = 0.398$ mas and $\theta_K = 0.401$ mas, 
HD 215144 with $\theta_H = 0.383$ mas  and $\theta_K = 0.384$ mas, 
HD 218918 with $\theta_H = 0.384$ mas  and $\theta_K = 0.385$ mas, and
HD 1367   with $\theta_H = 0.636 $ mas  and $\theta_K = 0.641$ mas. We used a conservative 5\% uncertainty on all calibrator angular diameters. Finally, the wavelengths in the reduced files were then divided by correction factors of $1.0054 \pm 0.002$ for MIRC-X and $1.0067 \pm 0.002$ for MYSTIC (J. D. Monnier, private communication) as suggested in the reduction manual.

We fit the relative position of the secondary components to the visibilities and closure phases from  MIRC-X and MYSTIC simultaneously using the method of \citet{schaefer16}\footnote{\href{http://www.chara.gsu.edu/analysis-software/binary-grid-search}{http://www.chara.gsu.edu/analysis-software/binary-grid-search}}, which searches across a grid of separations in right ascension and declination to find the best-fit relative position. Within each grid box, we compared the observed $V^2$ and CP to model values to fit for the relative position and flux ratio of the components with a least-squares optimizer, then calculated the reduced $\chi^2$ statistic. The individual component stars are unresolved in both of our binary systems. Because the component stars of both systems are very similar in temperature, their $H-$ and $K$-band flux ratios would also be the same, so we fit one flux ratio value to both the MIRC-X and MYSTIC data in order to fit the datasets simultaneously. Figure~\ref{vis2cp} in the Appendix shows an example of the fit to one night of observations. We then searched a small area around the best-fit position to find the contour marking $\chi^2 \le \chi^2_{min}+1$ that determines the major axis ($\sigma_{maj}$), minor axis ($\sigma_{min}$), and position angle ($\phi$) of each error ellipse.  Because the orbital periods of these systems are much longer than the observation time, any orbital motion occurring within a single night's observation is very small and contained within the error ellipse. The best-fit separations ($\rho$), position angles ($\theta$, measured East of North), error ellipse parameters, and average flux ratio estimates for each observation are listed in Table~\ref{relpos1} for HD~210763 and Table~\ref{relpos2} for HD~221950.  The flux ratio for $H$-band specifically can be found from the two nights with only MIRC-X data, which is 0.38 for HD~210763  and 0.71 for HD~221950; we used this value later in the SED analysis.

\begin{deluxetable*}{lccccccc}	
\tablewidth{0pt}
\tabletypesize{\footnotesize}
\tablecaption{  Relative Astrometry Measurements for HD 210763 \vspace{6pt} \label{relpos1}    }
\tablehead{ 
\colhead{UT Date} 
& \colhead{MJD} 
& \colhead{$\rho$ (mas)} 
& \colhead{$\theta$ (deg)} 
& \colhead{$\sigma_{maj}$ (mas)} 
& \colhead{$\sigma_{min}$ (mas)} 
& \colhead{$\phi$ (deg)} 
& \colhead{$f_2 / f_1$} 
}
\startdata
%    DATE	HJD		rho 	theta		dmaj 	dmin	 dphi	fr   
\hline
%  date			mjd				rho			theta		dmaj 		dmin		etheta  fr    
2020-08-12	&	59073.2774	&	1.834    &	305.3	&	0.011	&	0.007	&	45.0   &	0.38   \\
2020-08-13	&	59074.2737	&	1.757    &	314.8	&	0.009	&	0.008	&	30.0   &	0.38   \\
2024-08-04	&	60526.3730	&	2.932    &	398.9	&	0.009	&	0.008	&	5.0    &	0.38   \\
2024-08-07	&	60529.3626	&	3.437    &	407.9	&	0.028	&	0.018	&	35.0   &	0.38   \\
2024-08-11	&	60533.3682	&	3.956    &	416.5	&	0.024	&	0.017	&	170.0  &	0.36   \\
2024-08-13	&	60535.3252	&	4.104    &	420.8	&	0.043	&	0.033	&	5.0    &	0.37   \\
2024-08-20	&	60542.3624	&	3.525    &	433.7	&	0.028	&	0.018	&	110.0  &	0.36   \\
2024-08-28	&	60550.3488	&	1.936    &	260.4	&	0.019	&	0.013	&	340.0   &	0.38   \\
\enddata
\end{deluxetable*}

\begin{deluxetable*}{lccccccc}	
\tablewidth{0pt}
\tabletypesize{\footnotesize}
\tablecaption{  Relative Astrometry Measurements for HD 221950 \vspace{6pt} \label{relpos2}    }
\tablehead{ 
\colhead{UT Date} 
& \colhead{MJD} 
& \colhead{$\rho$ (mas)} 
& \colhead{$\theta$ (deg)} 
& \colhead{$\sigma_{maj}$ (mas)} 
& \colhead{$\sigma_{min}$ (mas)} 
& \colhead{$\phi$ (deg)} 
& \colhead{$f_2 / f_1$} 
}
\startdata
%    DATE	HJD		rho 	theta		dmaj 	dmin	 dphi	fr   instrument
\hline
%  date			mjd				rho			theta		dmaj 		dmin		etheta  fr    \\
2020-08-11 &	59072.3836	&  1.913 &  308.6 & 0.018 & 0.008 &  0.0	& 0.71 \\
2020-08-12 &	59073.3548	&  2.703 &  298.7 & 0.010 & 0.008 &  5.0	& 0.71 \\
2024-08-04 &	60526.4299	&  1.438 &  321.2 & 0.006 & 0.005 &  30.0	& 0.71 \\
2024-08-07 &	60529.4214	&  3.835 &  291.8 & 0.011 & 0.009 &  40.0	& 0.71 \\
2024-08-11 &	60533.4220	&  6.931 &  283.9 & 0.019 & 0.018 &  0.0	& 0.70 \\
2024-08-13 &	60535.3878	&  8.158 &  282.2 & 0.035 & 0.025 &  350.0	& 0.68 \\
2024-08-20 &	60542.4205	&  9.768 &  278.0 & 0.026 & 0.020 &  100.0	& 0.70 \\
2024-08-28 &	60550.4044	&  0.740 &  238.9 & 0.020 & 0.006 &  7.0	& 0.69 \\
\enddata
\end{deluxetable*}

\begin{figure*}[t!]
\centering
\includegraphics[width=0.49\textwidth]{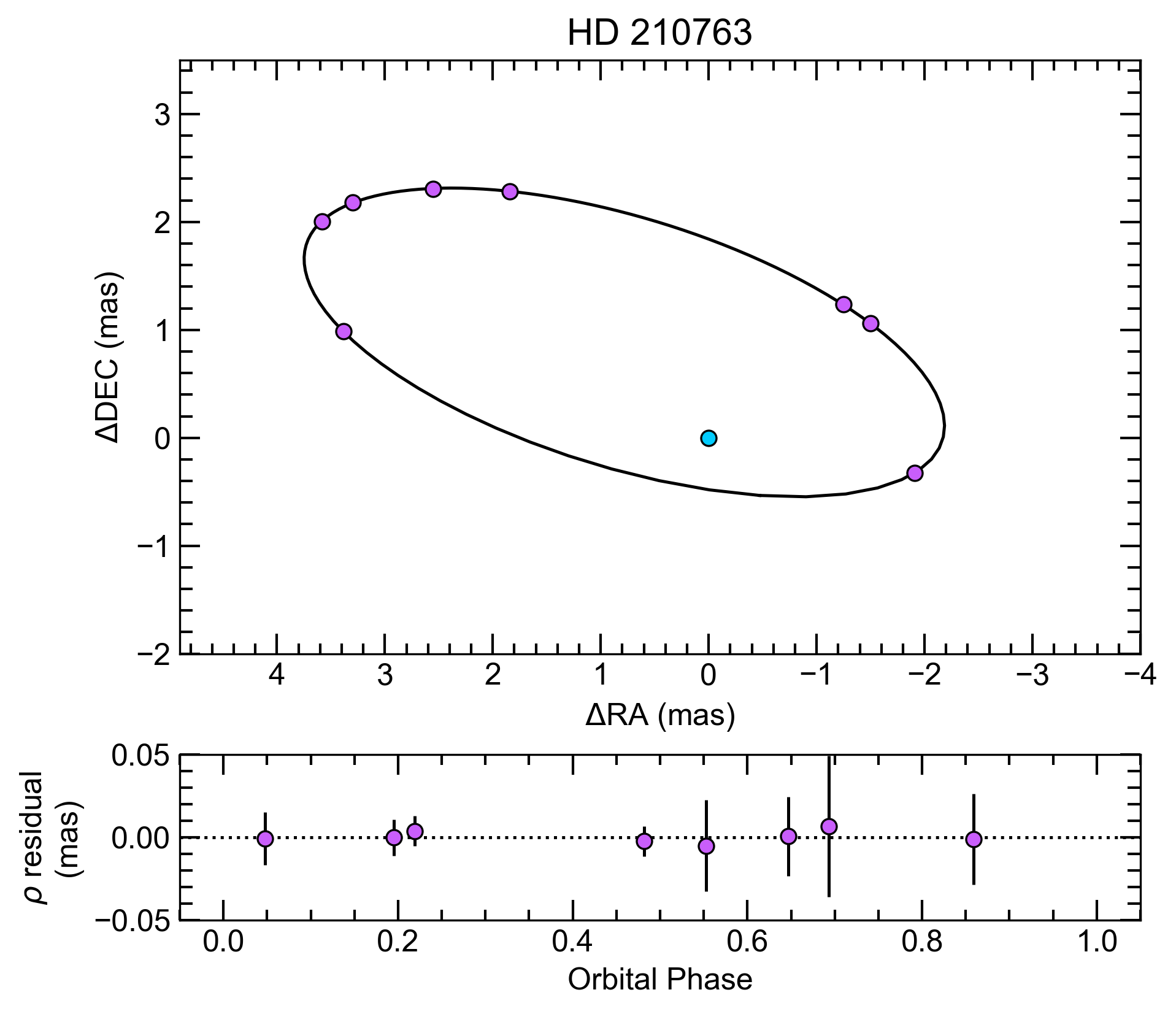}
\includegraphics[width=0.49\textwidth]{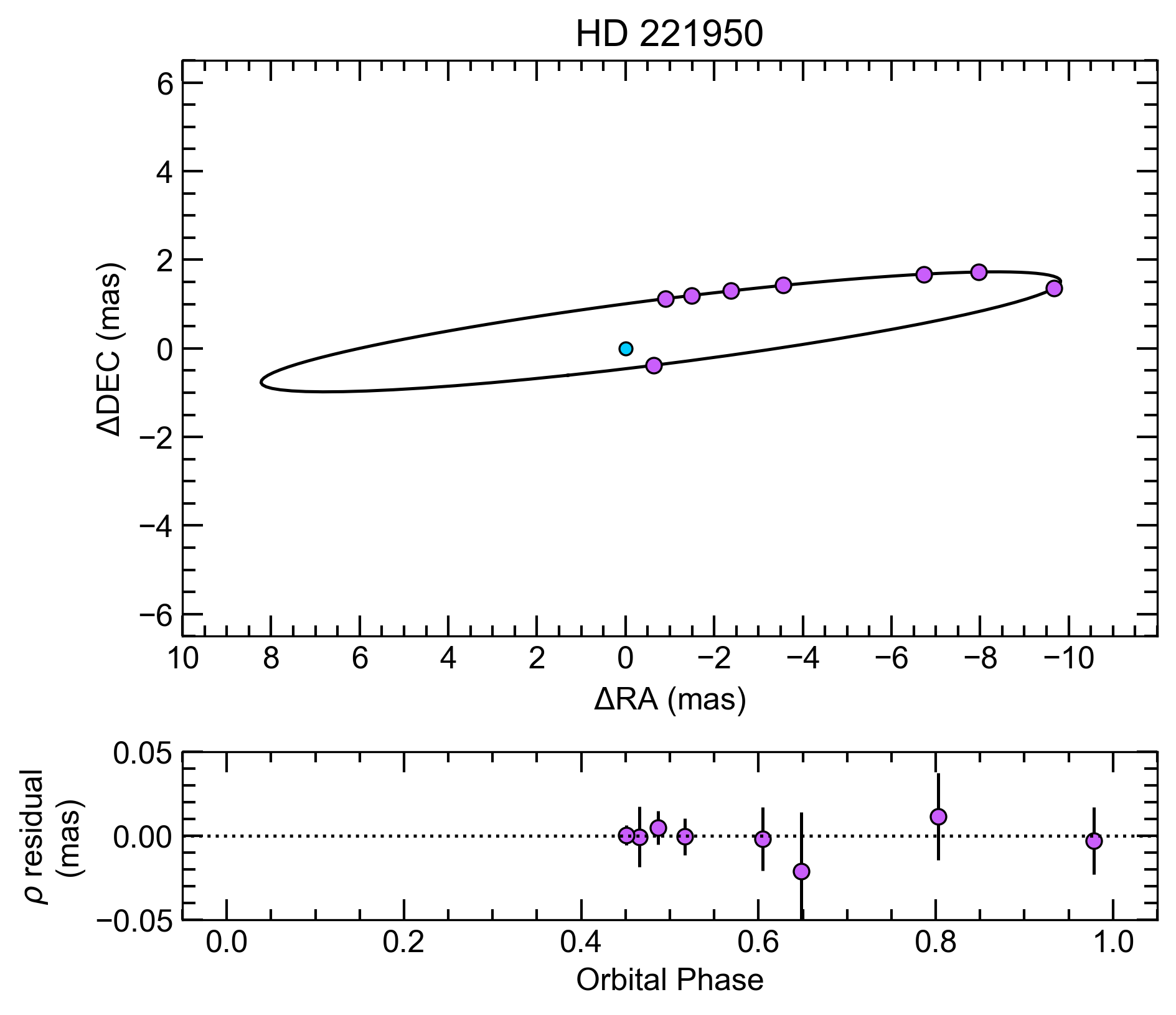}
\caption{Visual orbits of HD~210763 (left) and HD~221950 (right). The primary star is located at the origin (blue circle), and the relative positions of the secondary star are marked by purple circles. The error ellipses are all smaller than the size of the data points. The solid black curves represent the best-fit visual orbits. Residuals in the separation are shown in the bottom panels, with the error bars corresponding to the major axis of the error ellipse. 
\label{vborbit}}
\end{figure*}

\clearpage

\begin{figure*}
\centering
\includegraphics[width=0.49\textwidth]{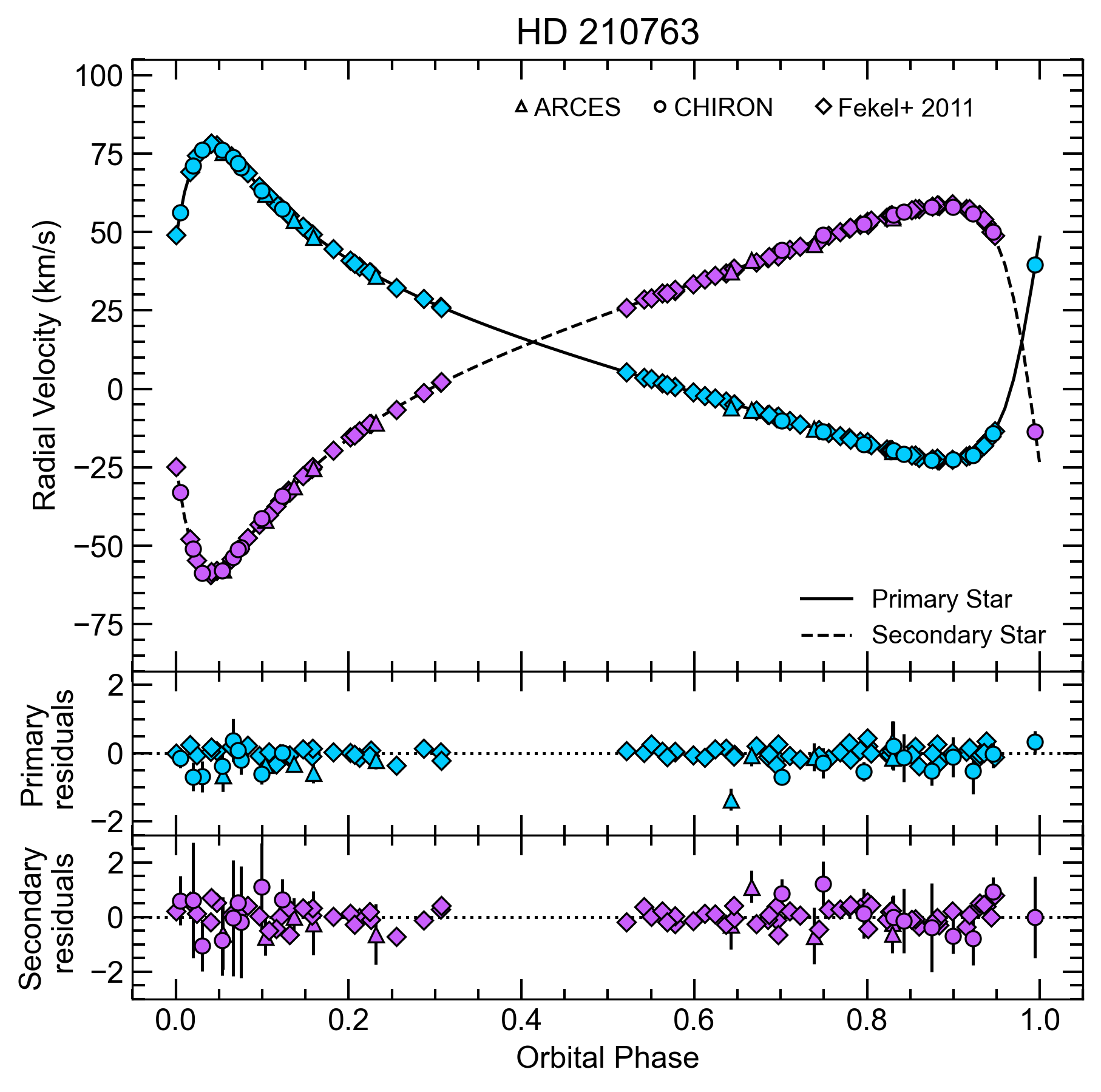}
\includegraphics[width=0.49\textwidth]{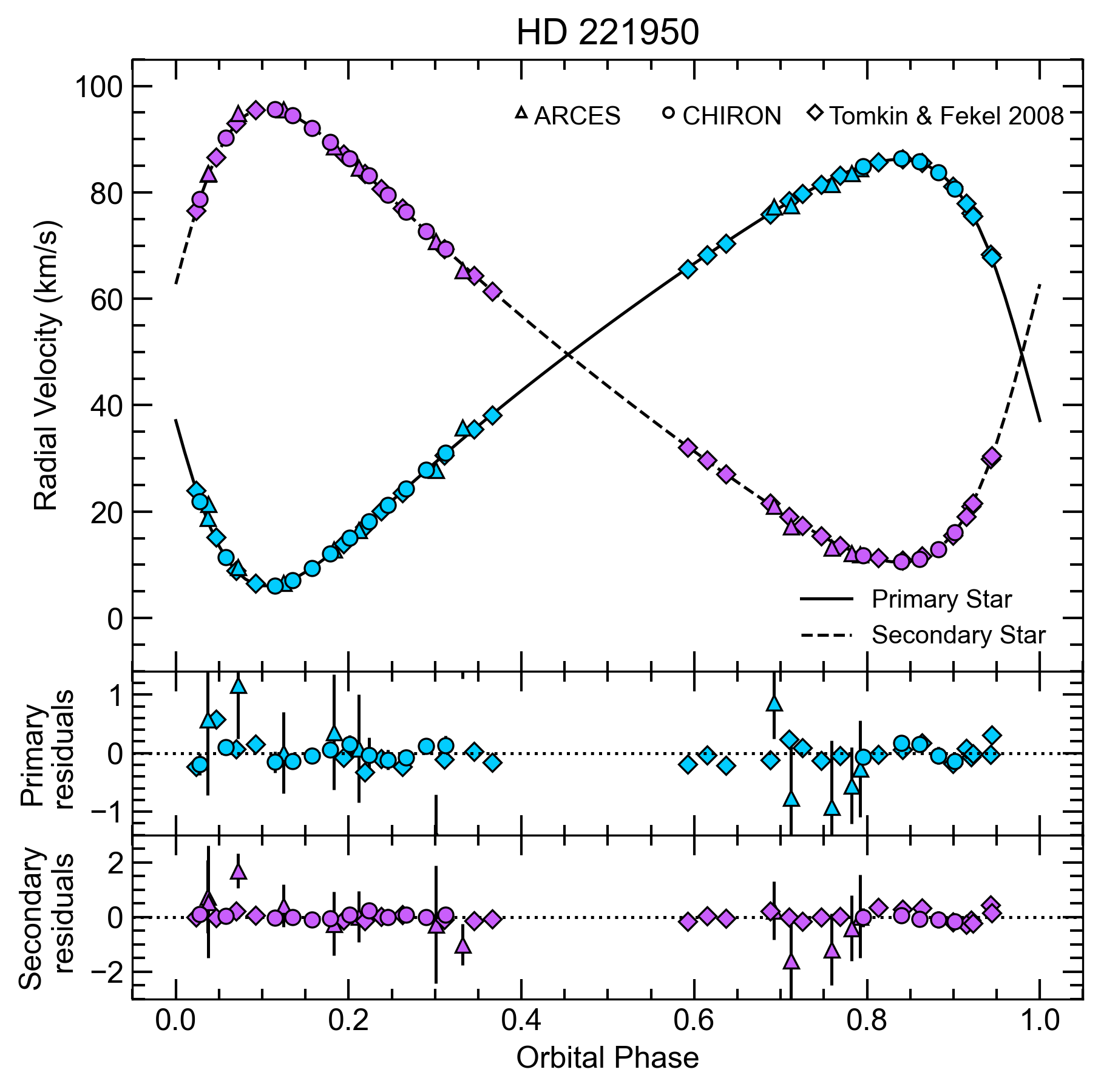}
\caption{Radial velocity curves of HD~210763 (left) and HD~221950 (right). The observed data for the primary and secondary star are shown with the blue and purple points, respectively. The triangles, circles, and diamonds represent the APO, CTIO, and literature data. The best-fit curves are shown as black lines, and the residuals to the fit are shown in the bottom panels.  
\label{rvcurve}}
\end{figure*}

% -----------------------------------------------------------------------------
\section{Orbital Solution}\label{sect:vbsbfit}

We determined the 3D orbital solutions by simultaneously fitting the interferometric and spectroscopic data using the method of \citet{schaefer16}\footnote{\href{https://www.chara.gsu.edu/analysis-software/orbfit-lib}{https://www.chara.gsu.edu/analysis-software/orbfit-lib}}. The full set of orbital parameters includes the orbital period ($P$), epoch of periastron ($T$), eccentricity ($e$), inclination ($i$, where $i=90^\circ$ corresponds to edge-on), longitude of periastron of the primary star ($\omega_1$), angular semi-major axis ($a$), longitude of the ascending node ($\Omega$), velocity semi-amplitudes ($K_1$, $K_2$), and  systemic velocity ($\gamma$).  The visual orbit solutions are shown in Figure~\ref{vborbit} and the spectroscopic orbit solutions are shown in Figure~\ref{rvcurve}. Table~\ref{orbpar} lists the best-fit orbital parameters and their $1\sigma$ uncertainties for each system. We used a bootstrapping analysis to determine the uncertainty of each orbital parameter, based on the standard deviation of the resulting distributions for each parameter. The corner plots showing these distributions are shown in Figures~\ref{corner1} and \ref{corner2} in the Appendix. Finally, we calculated the component star masses and binary distance from the combined orbital solution and bootstrapping uncertainties using nominal Solar values from \citet{prsa16}. 
HD~210763 has masses of 
$M_1 = 1.748 \pm 0.008 ~M_\odot$ and 
$M_2 = 1.493 \pm 0.006 ~M_\odot$, and  
HD~221950 has masses of 
$M_1 = 1.098 \pm 0.006 ~M_\odot$ and 
$M_2 = 1.031 \pm 0.005 ~M_\odot$. 
Our results for HD~210763 are consistent with those of \citet{gallenne23}, proving good agreement and similar precision between the CHARA and VLTI observations.

\begin{deluxetable}{lcc}
\tabletypesize{\footnotesize}
\tablecaption{Orbital \& Stellar Parameters of HD 210763 and HD 221950 \vspace{6pt} \label{orbpar}}
\tablehead{
\colhead{Parameter}       & \colhead{HD 210763}             & \colhead{HD 221950} } 
\startdata	                                       
$P$ (days)	          	& $42.38113	\pm 0.00008  $ 		& $45.45950 \pm 0.00008 $ \\
$T$ (MJD)				& $57708.7995\pm 0.0072  $ 		& $53323.3255 \pm 0.0094 $\\
$e$			          	& $0.6233 \pm 0.0005   $    	& $0.3802 \pm 0.0006 $	  \\
$\alpha$ (mas)			& $3.738 \pm 0.006   $     		& $9.79   \pm 0.02 $	  \\
$i$ (deg)				& $70.92\pm 0.13   $   	   		& $94.30  \pm 0.02 $	  \\
$\Omega$ (deg)		    & $77.58\pm 0.14   $   	   		& $277.07 \pm 0.08 $	  \\
$\omega_1$ (deg)	  	& $294.15\pm 0.06   $      		& $102.98 \pm 0.06 $	  \\
$K_1$ (km s$^{-1}$)	  	& $50.32\pm 0.03   $   	   		& $40.08 \pm 0.04 $	  	\\
$K_2$ (km s$^{-1}$)	  	& $58.90\pm 0.07   $   	   		& $42.66 \pm 0.04 $	  	\\
$\gamma$ (km s$^{-1}$)	& $14.90\pm 0.02   $   	   		& $49.52 \pm 0.02 $	  	\\
\hline                                             		
$M_1$ ($M_\odot$)	  	& $1.748 \pm 0.008 $   	   		& $ 1.098 \pm 0.006 $	  \\
$M_2$ ($M_\odot$)	  	& $1.493 \pm 0.006 $   	   		& $ 1.031 \pm 0.005 $	  \\
$R_1$ ($R_\odot$)	  	& $2.96 \pm 0.11 $   			& $1.30 \pm 0.08 $	  \\
$R_2$ ($R_\odot$)	  	& $1.81 \pm 0.07 $   			& $1.09 \pm 0.07 $	  \\
$T_{\rm eff~1}$ (K)	  	& $6470 \pm 100 $   			& $ 6390 \pm 130 $	  \\
$T_{\rm eff~2}$ (K)	  	& $6370 \pm 110 $   			& $ 6330 \pm 130 $	  \\
$a$ ($AU$)  		    & $0.352 \pm 0.0004 $      		& $ 0.321 \pm 0.0002$	  \\
$d$ (pc)  				& $93.62 \pm 0.21 $   	   		& $ 32.75 \pm 0.07$	  \\
\enddata                   
\end{deluxetable}

Double-lined visual binaries provide model-independent distances, which are useful tests of other measurement techniques like trigonometric parallax. We found the distance of 
HD~210763 to be 
$94.19\pm 0.21$ pc from orbital parallax, which is consistent with the distance of 
$93.81\pm0.26$~pc from the Gaia DR3 trigonometric parallax \citep{gaia1, gaia3}. 
For HD~221950, our distance from orbital parallax of 
$32.75 \pm 0.07$ pc is consistent with Gaia's distance of 
$32.55\pm0.66$~pc and improves the precision by an order of magnitude. Note, the Gaia distances include the suggested parallax corrections from \citet{lindegren21}\footnote{\href{https://gitlab.com/icc-ub/public/gaiadr3_zeropoint}{https://gitlab.com/icc-ub/public/gaiadr3\_zeropoint}}. 

\begin{figure}[h!]
\centering
\includegraphics[width=0.5\textwidth]{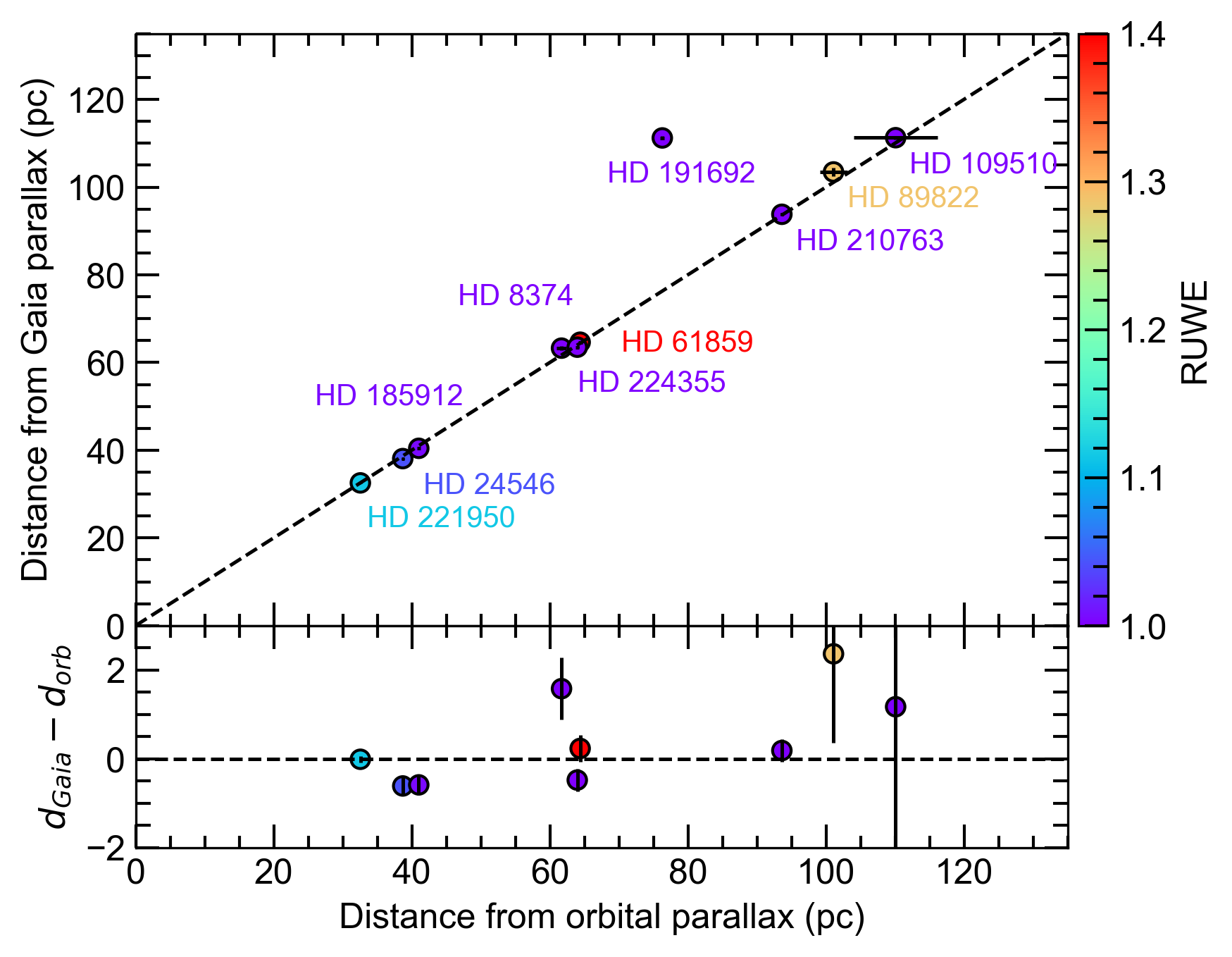}
\caption{Comparison of distances measured from orbital parallax (Papers I-V) and trigonometric parallax from Gaia DR3. The symbol color represents the RUWE value from Gaia. The bottom panel shows the residuals ($d_{Gaia} - d_{orb}$) with the larger of the distance uncertainties for each binary.  Some systems with a low RUWE value show a mismatch between the two distance methods, so RUWE is not an adequate indicator of the parallax accuracy. 
\label{distances}}
\end{figure}

For a binary system, the added flux of the second star and the orbital motion of the photo-center could bias the trigonometric parallax measurement. The commonly used indicator of this bias in Gaia data is the renormalized unit weight error (RUWE), where a value above 1.4 indicates a poor astrometric fit. Figure~\ref{distances} compares the distances from orbital and trigonometric parallax from all ten of the double-lined visual binaries from this paper series.  The systems with a higher RUWE do show inconsistency between the two methods as expected, but several of our binaries have a low RUWE value and show a mismatch in the distance measurements.  
\citet{stassun21} found a similar result using eclipsing binaries and cautioned that stars with RUWE values between 1 - 1.4 are often binary systems; therefore, the RUWE value alone is not an adequate indicator of the parallax accuracy nor of binarity. 

% only two of these systems have solutions in the non-single star catalog (HD 24546 and HD 224355); the other eight systems do not have orbital solutions in DR3. 

% -----------------------------------------------------------------------------
\section{Stellar Parameters}\label{sect:params}

\subsection{Effective Temperatures}\label{sect:specmatch}

We disentangled the observed CHIRON spectra of each binary into the individual component spectra using the Shift \& Add code \citep{shiftadd1, shiftadd2}, the $V$-band flux ratios from TODCOR, and the best-fit orbital parameters. We then compared the disentangled spectrum of each component star to the empirical library of stellar spectra in order to estimate the effective temperature and iron abundance with the \texttt{SpecMatch-emp} code \citep{specmatch}\footnote{\href{https://specmatch-emp.readthedocs.io/en/latest/}{https://specmatch-emp.readthedocs.io/en/latest/}}.  We found the best-fit temperature and $\feh$ for each of the 15 echelle orders between 5000--6000 \AA, then averaged the results for each component. 
HD~210763 has 
\teff$_1 = 6470 \pm 100$~K,   
\teff$_2 = 6370 \pm 110$~K, and 
$\feh = 0.15 \pm 0.10$. This is consistent with the estimates from \citet{fekel11} within the uncertainties, but \citet{gallenne23} measured a lower temperature for the primary component (6170 K) and a higher temperature for the secondary component (6520 K), possibly due to using a different metallicity (0.01 dex).
HD~221950 has 
\teff$_1 = 6390 \pm 130$~K,   
\teff$_2 = 6330 \pm 130$~K, and 
$\feh = -0.50 \pm 0.10$, also consistent with the estimates of \citet{tomkin08} within the uncertainties.

\begin{figure*}[t!]
\centering
\includegraphics[width=0.49\textwidth]{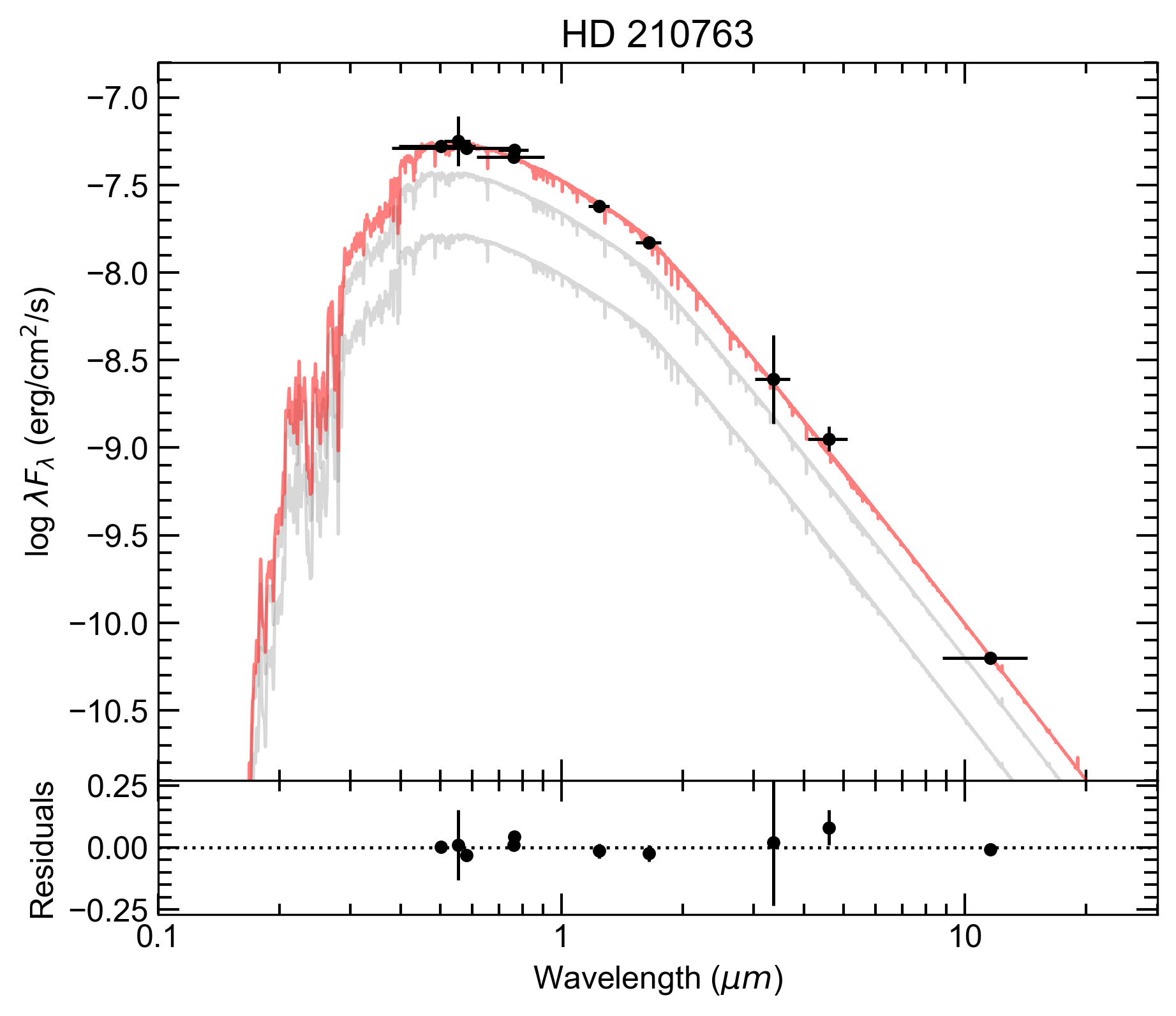}
\includegraphics[width=0.49\textwidth]{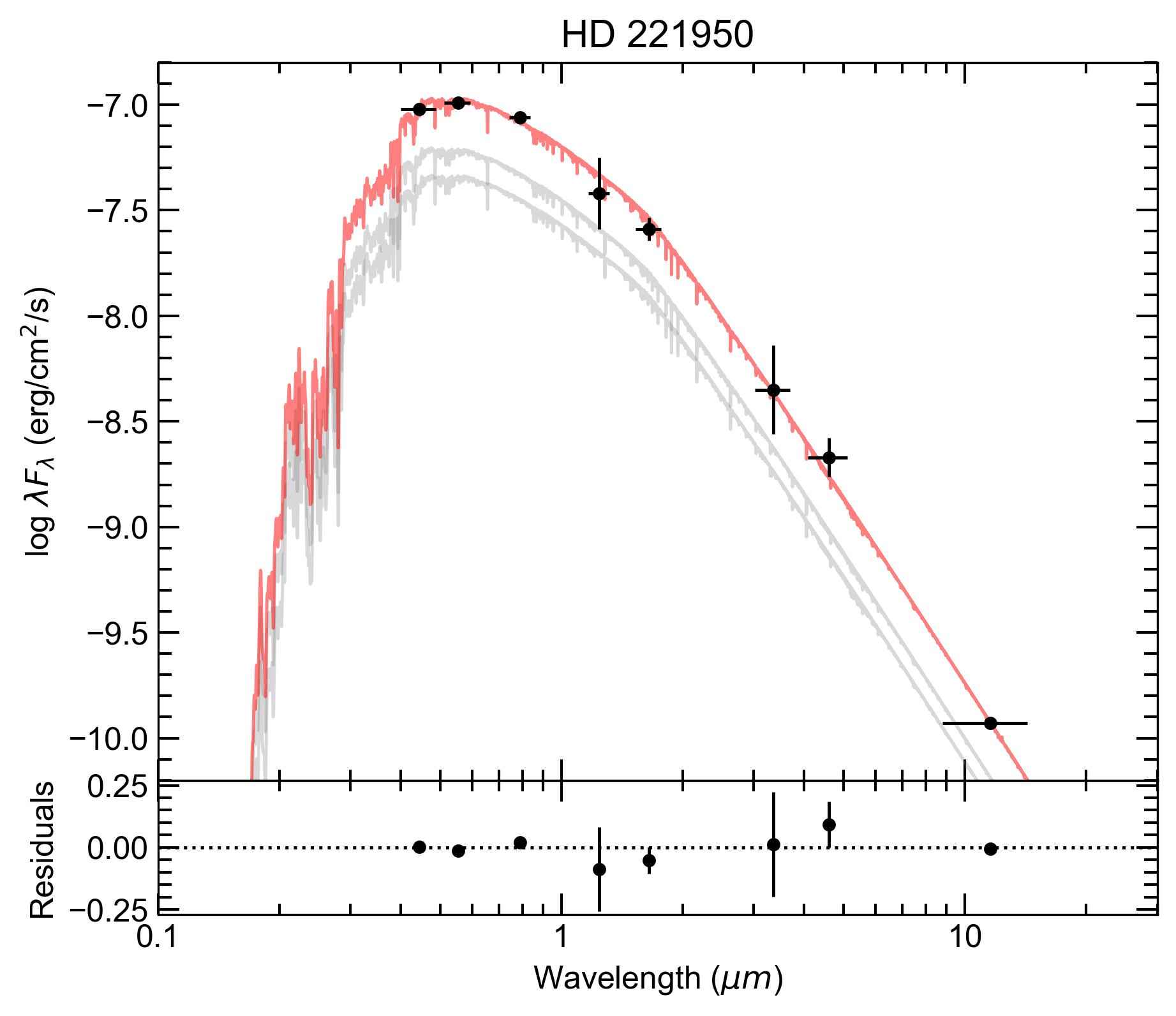}
\caption{Spectral energy distributions of HD~210763 (left) and HD~221950 (right). The observed fluxes are shown with black points, the best-fit binary model is shown as the red line, and the residuals to the fit are shown in the bottom panel. The SED models for the individual component stars are shown in grey.  
\label{sed}}
\end{figure*}

\subsection{Stellar Radii}\label{sect:sedfit}

We fit for the radius of each component star using a spectral energy distribution (SED) analysis and the observed fluxes of each binary system from UCAC4 \citep{ucac4}, Gaia DR3 \citep{gaia3}, SDSS \citep{sdss}, 2MASS \citep{2mass}, and WISE \citep{wise} archival photometry. We created a \texttt{BT-Settl} SED model \citep{btsettl1, btsettl2} of each component star using the temperatures found above and surface gravity estimates from the literature \citep{tomkin08, fekel11}. The binary's combined flux can be calculated from the surface fluxes of each component star ($F_1, F_2$), the radii ($R_1, R_2$), and the distance ($d$) to the binary:
$$F_{bin} = \frac{R_1^{~2} F_1 + R_2^{~2} F_2}{d^2}$$
The binary's combined flux was then corrected for reddening, using a chosen color excess ($E(B-V)$) value and the reddening curve of the Milky Way \citep{reddening}. We constrained the radius ratio ($R_2/R_1$) of each binary using the $H$-band flux ratio determined from the CHARA astrometry, then fit for the radius of the primary star and the color excess for each system with a least-squares optimizer. The best-fit SED models are shown in Figure~\ref{sed}. 
We found HD~210763 to have 
$R_1 = 2.93 \pm 0.13~R_\odot$, 
$R_2 = 1.80 \pm 0.08~R_\odot$, and 
$E(B-V) = 0.05 \pm 0.02$ mag.
This corresponds to bolometric luminosities of
$L_1 = 13.2 \pm 2.0 ~L_\odot$ and 
$L_2 = 5.3 \pm 0.8 ~L_\odot$.
We found HD~221950 to have 
$R_1 = 1.30 \pm 0.08~R_\odot$, 
$R_2 = 1.09 \pm 0.07~R_\odot$, and 
$E(B-V) = 0.02 \pm 0.01$ mag. The resulting bolometric luminosities are then 
$L_1 = 2.5 \pm0.4 ~L_\odot$ and 
$L_2 = 1.7 \pm0.3 ~L_\odot$.

% -----------------------------------------------------------------------------

\begin{figure*}[t!]
\centering
\includegraphics[width=0.49\textwidth]{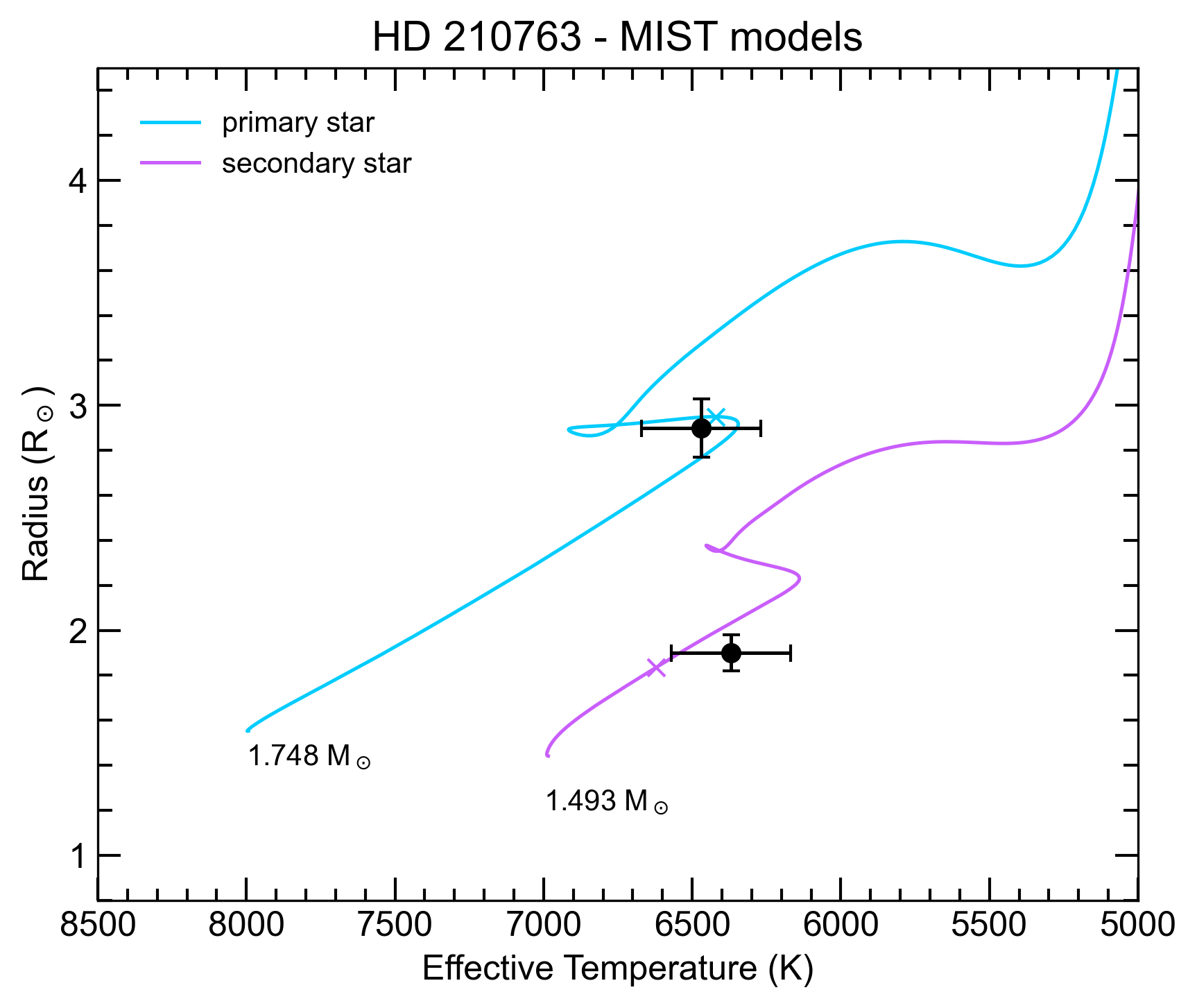}
\includegraphics[width=0.49\textwidth]{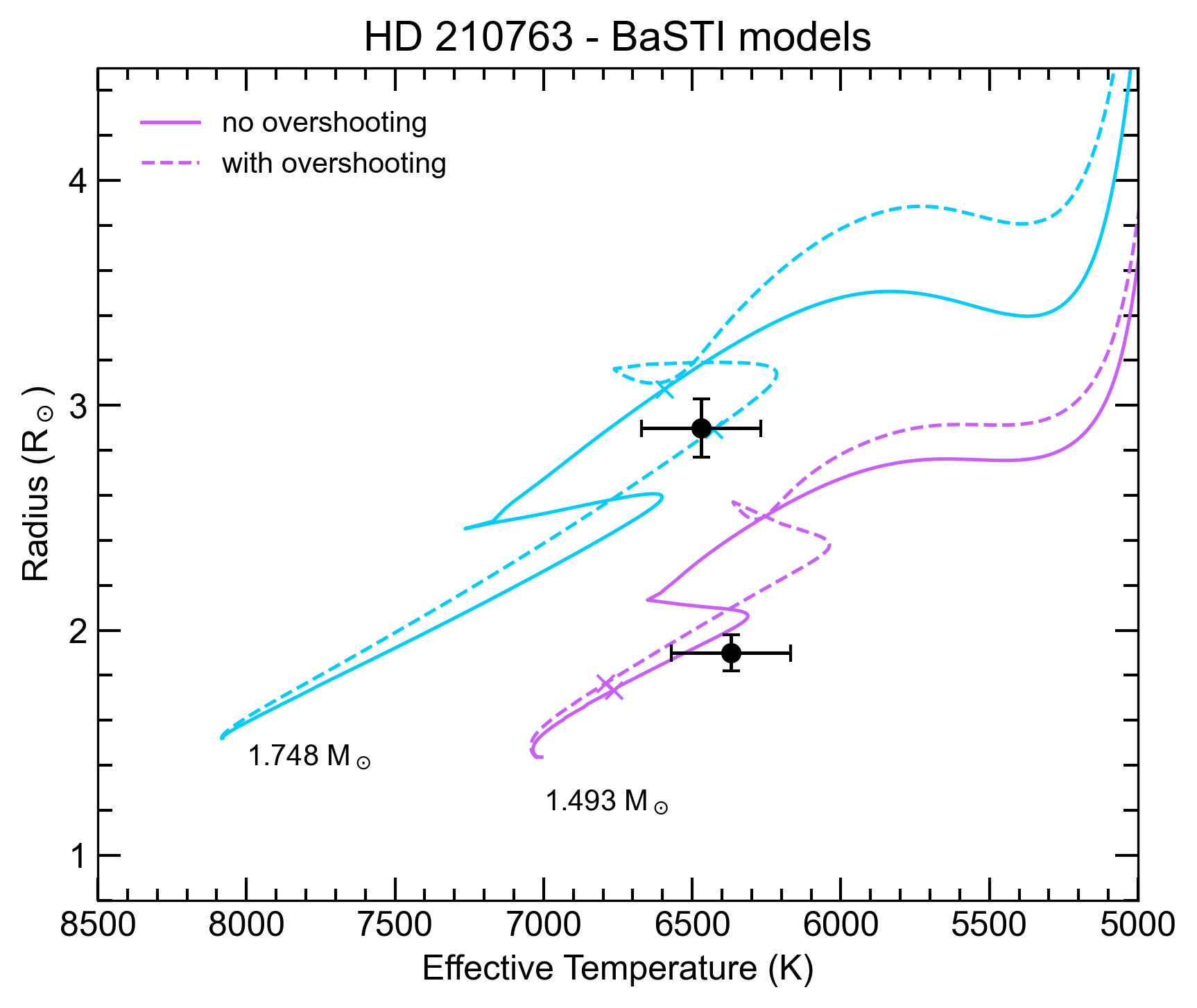}
\caption{Evolutionary tracks for HD~210763 using the MIST models (left) and BaSTI models (right). The observed temperatures and radii are shown as black points. The model evolutionary tracks are plotted in blue and purple for the primary and secondary stars, respectively.  An ``x" marks each evolutionary track at the primary star's best-fit age of 1.60 Gyr for the MIST model, 1.67 Gyr for the BasTI model without overshooting,  and 1.50 Gyr for the BasTI model with overshooting.
\label{evo1}}
\end{figure*}

\begin{figure*}[t!]
\centering
\includegraphics[height=7cm]{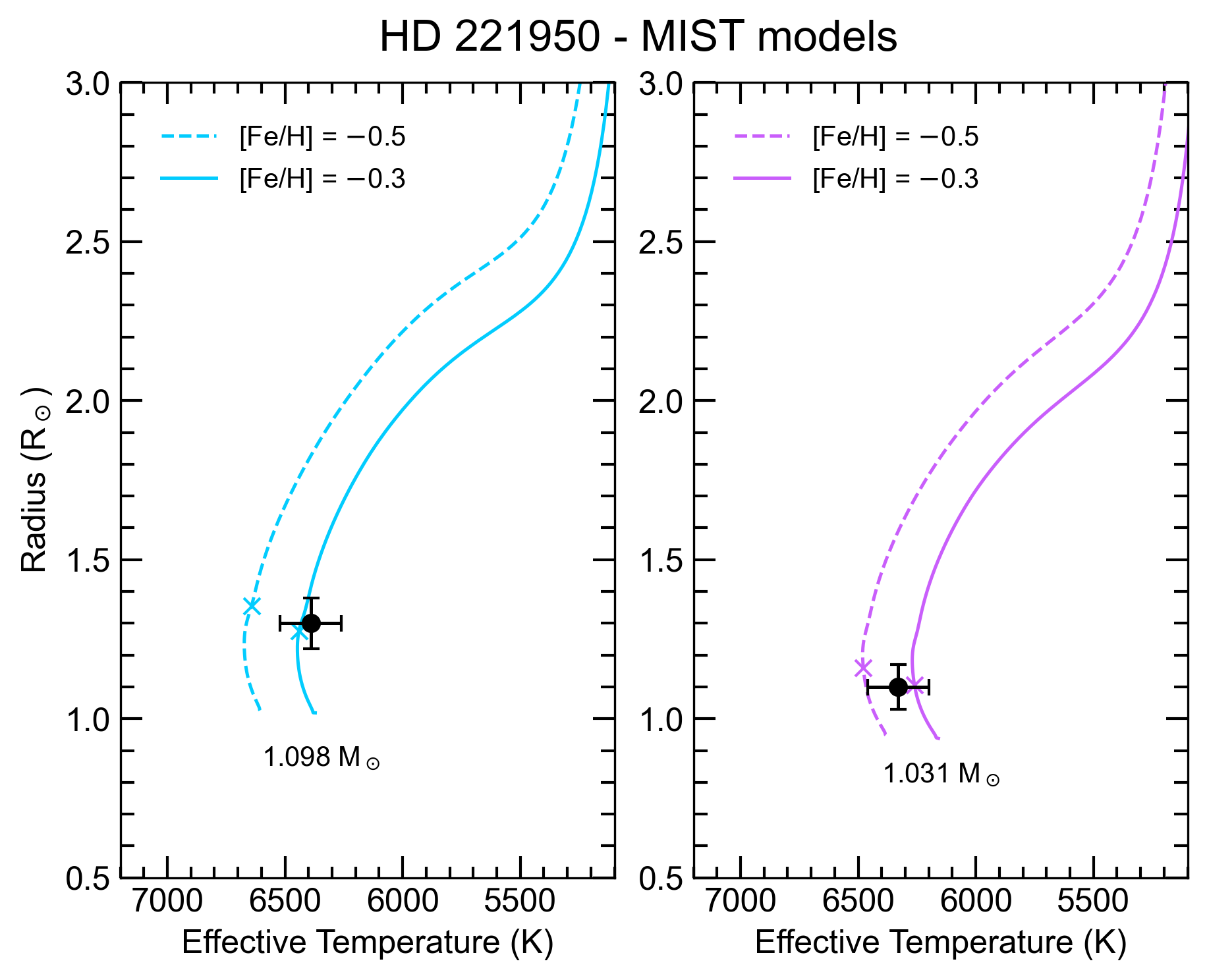}
\includegraphics[height=7cm]{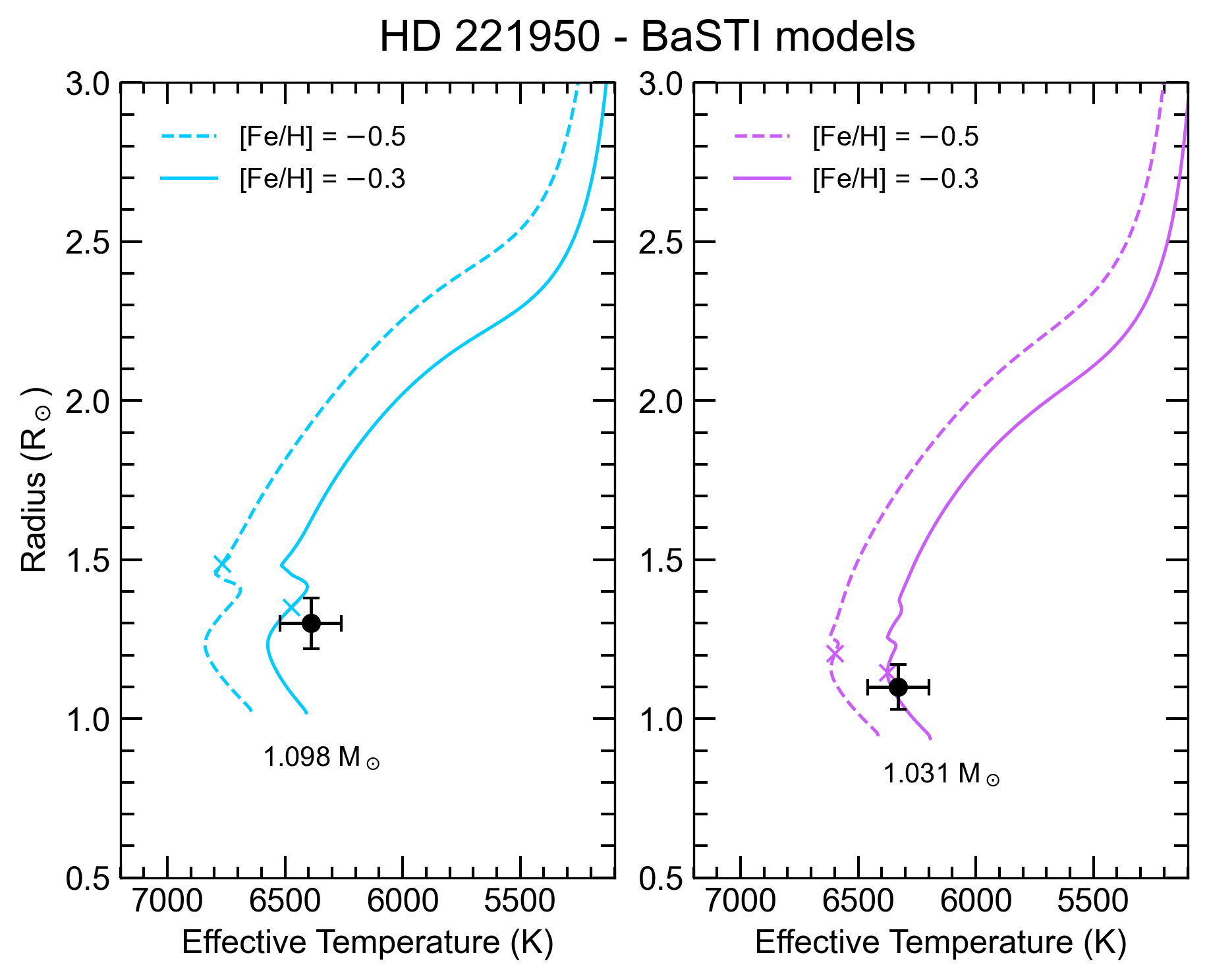}
\caption{Evolutionary tracks for HD~221950 using the MIST models (left) and BaSTI models (right). The observed temperatures and radii are shown as black points. The model evolutionary tracks are plotted in blue and purple for the primary and secondary stars, respectively.  An ``x" marks each evolutionary track at the primary star's best-fit age of 3.5 Gyr for the MIST models and 4.4 Gyr for the BasTI models.
\label{evo2}}
\end{figure*}

\section{Comparison with Evolutionary Models}\label{sect:evofit}

We compared the derived stellar parameters to evolutionary models in order to estimate the age of each binary and test the accuracy of these models. We tested the MIST models \citep[v2.5,][]{mist1, mist2} and the BaSTI-IAC models \citep{basti}, which are both non-rotating models with scaled-Solar abundances. For each binary, we used their online interpolators\footnote{\href{https://mist.science/interp_tracks.html}{https://mist.science/interp\_tracks.html}}\footnote{\href{http://basti-iac.oa-abruzzo.inaf.it/tracks.html}{http://basti-iac.oa-abruzzo.inaf.it/tracks.html}} to create custom evolutionary tracks for the masses found in Section~\ref{sect:vbsbfit} and iron abundances found in Section~\ref{sect:specmatch}.  

\textit{HD 210763 -- }
The evolutionary tracks for HD~210763 are shown in Figure~\ref{evo1}. For the MIST models, the primary star's parameters matched the evolutionary track at ages between 1.5--1.7 Gyr (within the uncertainties), while the secondary star matched  between 1.0--2.0 Gyr.  The primary star evolves faster and is right at the end of the main sequence, so it provides a tighter age constraint at $1.6 \pm 0.1$ Gyr. The secondary star's observed parameters are slightly outside the model track at this age, but within the $2\sigma$ uncertainties. 
For the BaSTI models, we tested models with and without convective core overshooting. The models with overshooting matched the primary star at 1.6--1.7 Gyr, but matched the secondary star at 2.1--2.2 Gyr. The models without overshooting matched the primary star at only 1.50 Gyr but were farther away from its observed data point than the models with overshooting. They matched the secondary star at 1.9--2.1 Gyr. Therefore, neither pair of BaSTI models were able to match the observed stellar parameters at the same age. Because of this, we adopt the MIST value of $1.6 \pm 0.1$ Gyr as the age of the system. This result is consistent with the estimate from \citet{gallenne23}.

\textit{HD 221950 -- }
The evolutionary tracks for HD~221950 are shown in Figure~\ref{evo2}. For the MIST models, we first tried the iron abundance of [Fe/H]$=-0.5$ as found in Section~\ref{sect:specmatch}, but these evolutionary tracks are not consistent with the observed parameters. We then tested [Fe/H]$=-0.3$, which provided slightly cooler models and successfully matched the observed parameters.  The primary star matched these evolutionary tracks at ages 3.0--4.1 Gyr and the secondary star matched between 2.2--4.4 Gyr, so the average system age is  $3.5 \pm 0.5$ Gyr according to the MIST models. 
For the BaSTI models, the tracks with and without convective core overshooting were indistinguishable, so we used only the models with overshooting. These models had the same problem as the MIST models, where the \feh ~from SpecMatch did not provide a good fit (shown in Figure~\ref{evo2}). Models with \feh$=-0.3$ provided a better fit, matching the primary star between 4.2--4.6 Gyr and the secondary star between 2.5--4.7 Gyr. Therefore, the BaSTI models were able to match the observed parameters at an average system age of $4.4 \pm 0.2$ Gyr.   
Combining the age results for the MIST and BaSTI models, the weighted average age for this system is then $3.76 \pm0.38$ Gyr.

% -----------------------------------------------------------------------------
\section{Conclusions}\label{sect:discussion}

By combining astrometric measurements from long-baseline optical interferometry and radial velocities from echelle spectroscopy, we measured the visual and spectroscopic orbits of two binary systems HD~210763 and HD~221950. We determined the masses of each component to within 0.5\% uncertainty and the distance to within 0.2\% uncertainty. Our masses are precise enough to use for testing stellar evolution models \citep{torres10}, but our radii do not yet have the necessary precision (6\% uncertainty). We estimate the component angular diameters to be about 0.15 and 0.09 mas for HD~210763 and 0.18 and 0.15 mas for HD 221950. Future work to directly measure the angular diameters will be possible with the longer telescope baselines in development at CHARA. Furthermore, our distances from orbital parallax provide model-independent confirmations of the Gaia distances, improving on the precision for HD 221950 by a factor of 10. 

We compared the observed stellar parameters to the predictions of the MIST and BaSTI stellar evolution models. For HD~210763, only the MIST models could match the observed parameters at the same age.  For HD~221950, both MIST and BaSTI models successfully matched the observed parameters at the same age, but the ages were inconsistent with each other. Both models also needed a slightly higher \feh~ value than what was measured from the disentangled spectra. MIST v2 adopts the \citet{gs98} solar abundance pattern, BaSTI uses the \citet{caffau11} pattern, and \texttt{SpecMatch-emp} sources metallicity values from the literature, so an inconsistency in how the abundances are defined could be causing this discrepancy. Binary stars also test the convective core overshooting prescriptions in stellar evolution models. MIST being able to match both binary systems provides evidence that the amount of convective core overshooting in their models is realistic. The BaSTI models that included overshooting did provide a better fit for the most massive star (HD 210763 A), but was an equally good fit for the other component stars as the models without overshooting. This is consistent with the results of \citet{ct18}, where the amount of overshooting increases with stellar mass.

% -----------------------------------------------------------------------------
\begin{acknowledgments}

\small{
The authors would like to thank the staff at APO, CHARA, and CTIO for their invaluable support during observations. 
%We would also like to thank the anonymous referee for their helpful comments. 
% MASGC
This work was supported by undergraduate fellowships from the Massachusetts Space Grant Consortium and the Mount Holyoke LYNK program.
% telescopes
This work is based upon observations obtained with 
the Apache Point Observatory 3.5-meter telescope, owned and operated by the Astrophysical Research Consortium;
the Georgia State University Center for High Angular Resolution Astronomy Array at Mount Wilson Observatory;
and the Cerro Tololo Inter-American Observatory 1.5m telescope.
% CHARA
The CHARA Array is supported by the National Science Foundation under Grant No. AST-2034336 and AST-2407956. Institutional support has been provided from the GSU College of Arts and Sciences, Office of the Provost, and Office of the Vice President for Research and Economic Development. 
Time at the CHARA Array was granted through the NOIRLab community access program (NOIRLab PropID: 2024B-854396; PI: K. Lester). 
% MIRCX
SK acknowledges funding for MIRC-X from the European Research Council under the European Union's Horizon 2020 research and innovation program (Starting Grant No. 639889 and Consolidated Grant No. 101003096). JDM acknowledges funding for the development of MIRC-X (NASA-XRP NNX16AD43G, NSF-AST 2009489) and MYSTIC (NSF-ATI 1506540, NSF-AST 1909165). 
%CHIRON
The CHIRON spectrograph on the Small and Moderate Aperture Research Telescope
System (SMARTS) 1.5 m telescope is operated as part of the SMARTS Observatory by REsearch Consortium On Nearby Stars (\href{http://www.recons.org/}{RECONS}) members. 
% other
This work has also made use of the Jean-Marie Mariotti Center Aspro \& SearchCal services, the CDS Astronomical Databases SIMBAD and VIZIER, and the Gaia archive. 
}

\end{acknowledgments}

\facilities{APO:3.5m, CHARA, CTIO:1.5m} 

\software{ 
AstroPy, NumPy, SciPy,
Grid Search for Binary Stars \citep{schaefer16}, 
MIST \citep{mist1, mist2}, 
SearchCal \citep{searchcal}, 
TODCOR \citep{todcor1}, 
}

% -----------------------------------------------------------------------------
\appendix

\begin{figure*}[h!]
\centering
\includegraphics[width=0.99\textwidth]{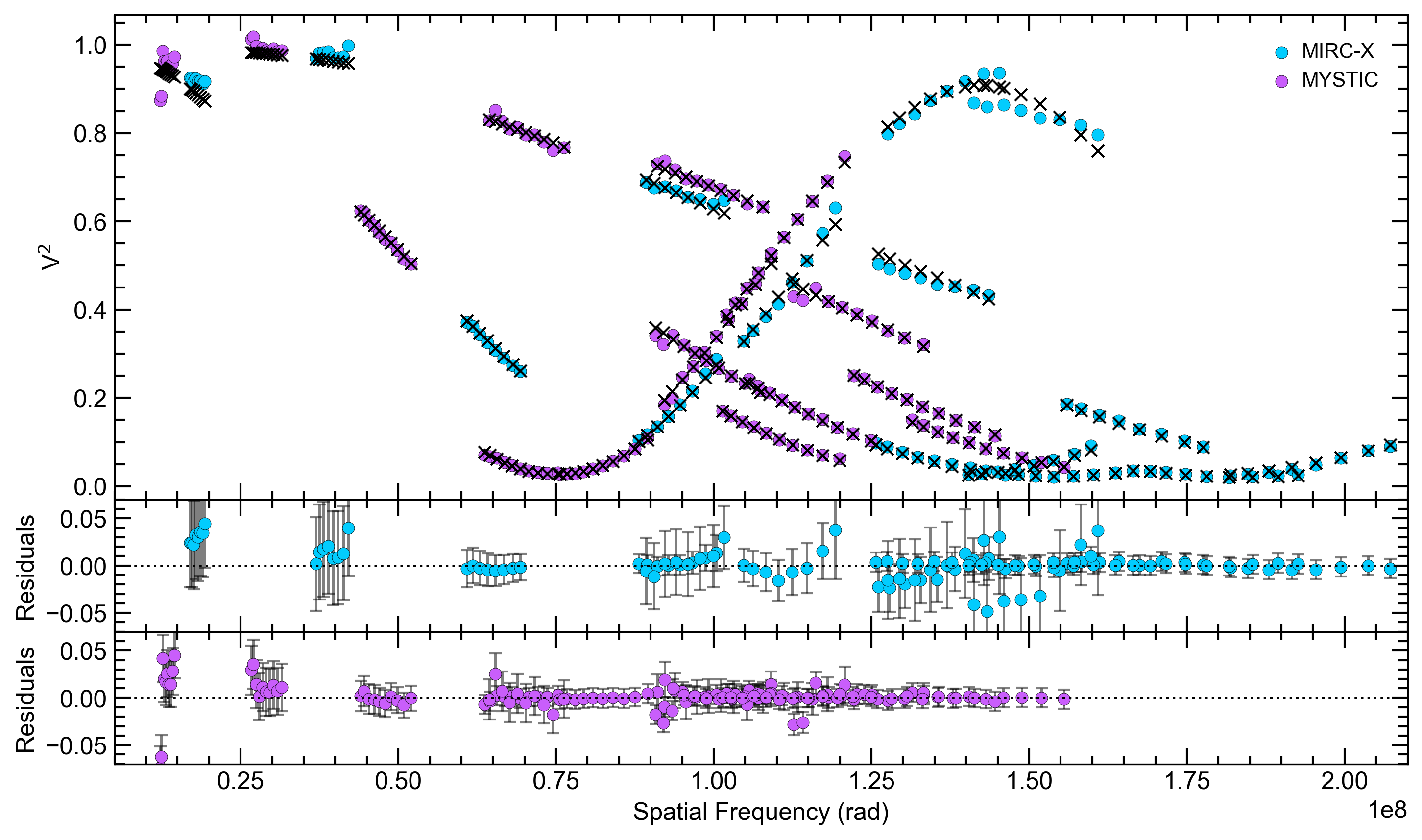}
\includegraphics[width=0.99\textwidth]{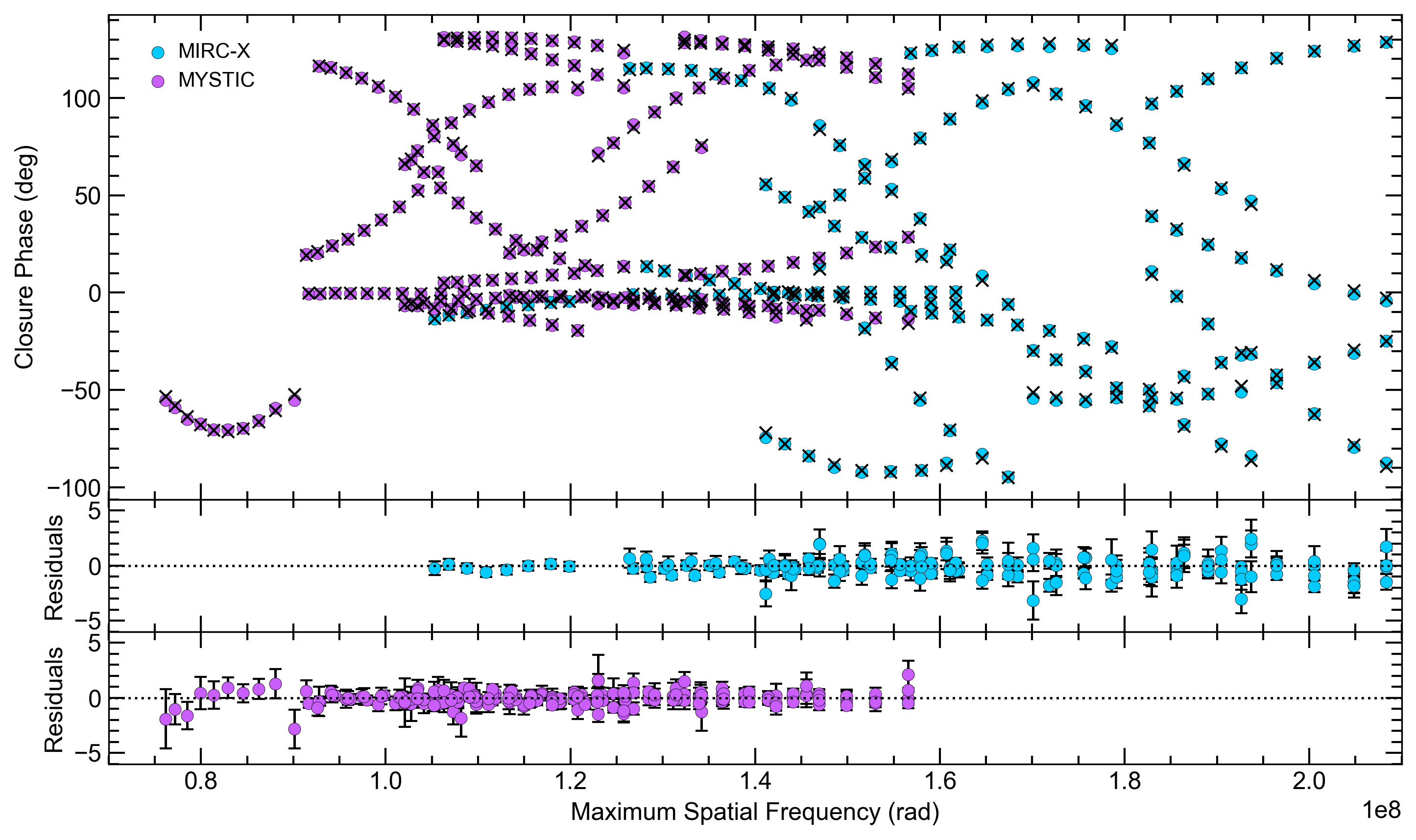}
\caption{Example CHARA observations of HD~221950 from 2024-08-04. The top plot shows the squared visibilities and the bottom plot shows the closure phases, where the MIRC-X data are in blue, the MYSTIC data are in purple, and the model values for the best-fit position are black X's. 
\label{vis2cp}}
\end{figure*}

\begin{figure*}[h!]
\centering
\includegraphics[width=0.99\textwidth]{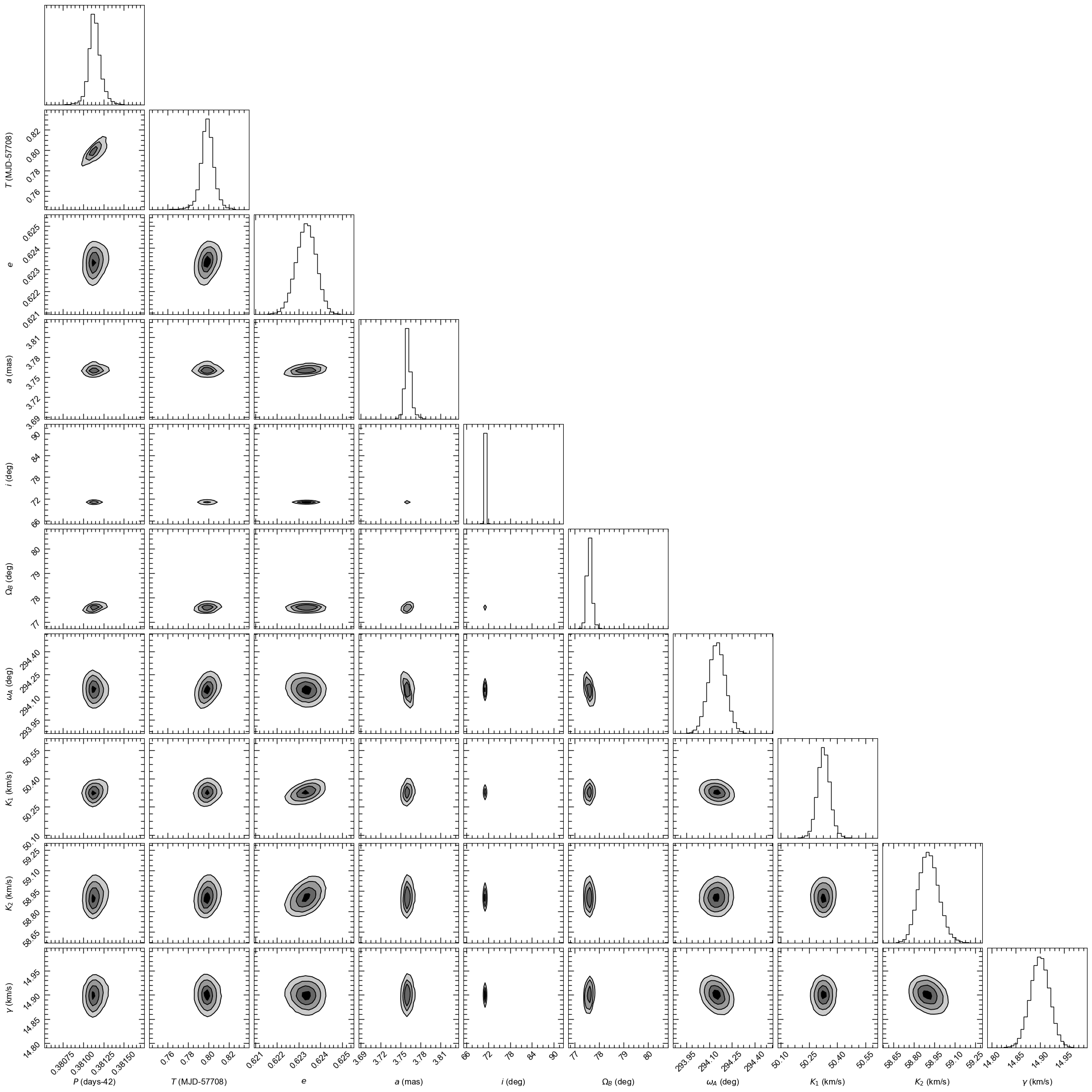}
\caption{Corner plot for HD~210763 with the results of our bootstrapping analysis to measure the uncertainties in each orbital parameter.
\label{corner1}}
\end{figure*}

\begin{figure*}[h!]
\centering
\includegraphics[width=0.99\textwidth]{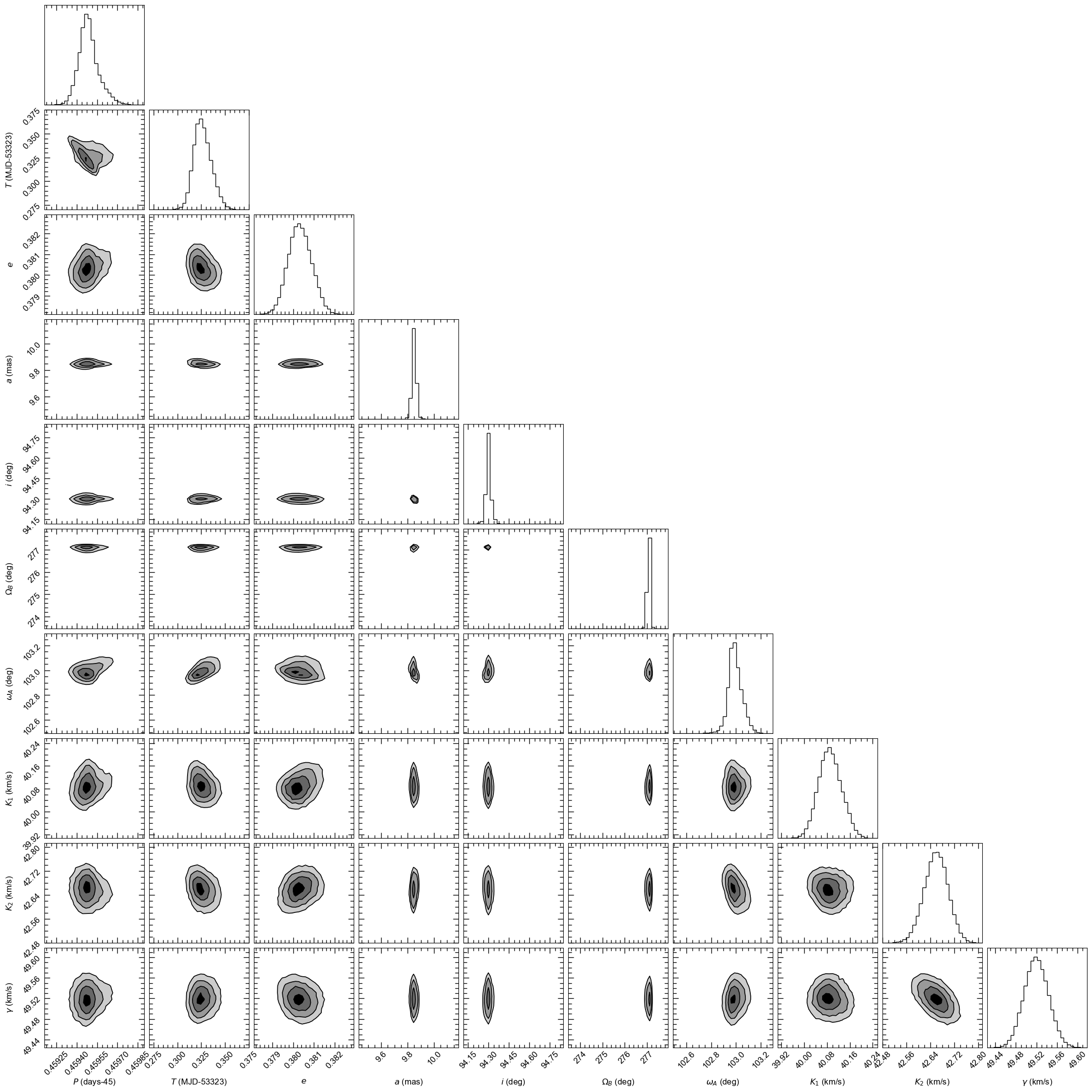}
\caption{Corner plot for HD~221950 with the results of our bootstrapping analysis to measure the uncertainties in each orbital parameter.
\label{corner2}}
\end{figure*}

Figure~\ref{vis2cp} shows an example of the relative astrometry fit to one night of CHARA observations. Figures \ref{corner1} and \ref{corner2} show the results of the bootstrapping error analysis for combined VB+SB2 orbit solution.

\clearpage

% -----------------------------------------------------------------------------
\bibliographystyle{aasjournal}       

\bibliography{references}

\end{document}